\documentclass[a4paper,fleqn,usenatbib]{mnras}

\usepackage{graphicx} 
\usepackage{txfonts} 
\usepackage{natbib} 
 
\newcommand{\Mpc}{h^{-1}\, {\rm Mpc}}

\newcommand{\be}{\begin{equation}}
\newcommand{\ee}{\end{equation}}
\newcommand{\bea}{\begin{equation}\begin{aligned}} 
\newcommand{\eea}{\end{aligned}\end{equation}}

\newcommand{\dd}[1]{{\rm d}#1\,}

\def\apj{ApJ} 
\def\apjl{ApJL} 
\def\apjs{ApJS} 
 
\def\aap{A\&A} 
\def\mnras{MNRAS}

 \newcommand{\IS}[1]{\textcolor{black}{#1}}
\newcommand{\ME}[1]{\textcolor{black}{#1}}
\title[Evolution of matter and galaxy clustering]{Evolution of matter and galaxy clustering in cosmological hydrodynamical simulations}
 
\author[J. Einasto et al.]{
Jaan Einasto,$^{1,2,3}$\thanks{E-mail: jaan.einasto@ut.ee} 
Gert H\"utsi,$^4$
Lauri-Juhan Liivam\"agi,$^{1}$
Changbom Park,$^{5}$\\
\newauthor
Juhan Kim, $^{5}$
Istvan Szapudi,$^6$
and Maret Einasto$^{1}$
\\
$^{1}$Tartu Observatory, 61602 T\~oravere, Estonia\\  
$^{2}$Estonian Academy of Sciences, 10130 Tallinn, Estonia\\
$^{3}$ICRANet, Piazza della Repubblica 10, 65122 Pescara, Italy \\ 
$^4$National Institute of Chemical Physics and Biophysics, 
Tallinn 10143, Estonia\\
$^5$Korea Institute for Advanced Study, 85 Hoegi-ro, Dongdaemun-gu,
Seoul 02455, Republic of Korea\\
$^6$Institute for Astronomy, University of Hawaii,
2680 Woodlawn Dr, Honolulu, HI 96822
} 

\date{Accepted 2023 June 1. Received 2023 June 1,  in origin form 2023 April 18; }  
\pubyear{2023}

\begin{document}
\label{firstpage}
\pagerange{\pageref{firstpage}--\pageref{lastpage}}
\maketitle
 
\begin{abstract}

  \IS{We quantify the evolution of matter and galaxy clustering in
    cosmological hydrodynamical simulations via correlation and bias
    functions of matter and galaxies. We use simulations TNG100 and
    TNG300 with epochs from $z=5$ to $z=0$. We calculate spatial
    correlation functions (CF) of galaxies, $\xi(r)$, for simulated
    galaxies and dark matter (DM) particles to characterise the
    evolving cosmic web. We find that bias parameters decrease during
    the evolution, confirming earlier results.  Bias parameters of the
    lowest luminosity galaxies, $b_0$, estimated from CFs are lower
    relative to CFs of particle density-limited clustered samples of
    DM.  At low and medium luminosities, bias parameters of galaxies
    are equal, suggesting that dwarf galaxies reside in the same
    filamentary web as brighter galaxies.  We find that bias
    parameters $b_0$, estimated from CFs of clustered DM, agree with
    the expected values from the fraction of particles in the
    clustered population, $b=1/F_c$.  The cosmic web contains
    filamentary structures of various densities, and fractions of
    matter in the clustered and the unclustered populations are both
    less than unity. Thus the CF amplitude of the clustered matter is
    always higher than for all matter, i.e. bias parameter must be
    $b>1$. Differences between CFs of galaxies and clustered DM
    suggest that these functions describe different properties of the
    cosmic web.}

\end{abstract}


\section{Introduction}

The clustering of galaxies and matter and their evolution are central
problems of cosmology.  Differences in the distribution of galaxies
and matter were noticed already in early studies by
\citet{Joeveer:1978dz, Joeveer:1978pb}, \citet{Gregory:1978} and
\citet{Tully:1978}, who showed that galaxies have filamentary
distribution with large regions devoid of galaxies.  Numerical
simulations by \citet{Doroshkevich:1980, Doroshkevich:1982fk}
predicted the filamentary character of particles but showed the
presence of a rarefied population of particles in voids.  This
difference was explained by \citet{Zeldovich:1982kl} as an indication
of a threshold mechanism in galaxy formation: galaxies form in
high-density filaments and knots but not in low-density regions --
cosmic voids. A physical mechanism for the formation of galaxies in
dark matter (DM) halos was suggested by \citet{White:1978}.  A more
detailed study of the formation of galaxies in the cosmic web was
presented by \citet{Dekel:1986aa, Dekel:1986ab, Dekel:1987aa} and
\citet{Bond:1996fv}.  Thus galaxies are biased tracers of matter
density fields.

Traditionally the difference between the distributions of galaxies and
matter is quantified by the bias parameter, which is defined
by the ratio of correlation functions or power spectra of galaxies and
matter, $b= \sqrt{\xi_g/\xi_m}$ \citep{Kaiser:1984}.  Most recent bias
studies are devoted to various aspects of the formation and evolution
of galaxies; for a review, see \citet{Desjacques:2018qf}.  This
paper focuses on the evolution of matter and galaxy
clustering.

Early studies showed that matter could be divided into two
populations: the clustered population with galaxies and systems of
galaxies, and the unclustered populations in low-density filaments and
voids. \IS{Unclustered in this context means much lower correlations
  than that of galaxies, but not necessarily zero correlations.}
\citet{Einasto:1987kw}  and \citet{Einasto:1994aa} identified the clustered matter
with samples of DM particles with local densities above a certain
threshold, $\rho \ge \rho_0$, and the unclustered matter with local
densities $\rho <\rho_0$.
\citet{Einasto:1994aa, Einasto:1999ku, Einasto:2023aa} investigated
the relation between clustered and total matter using simple analytic
models.  They found that the bias parameter $b$ is related to the
fraction of matter in the clustered population, $F_c$ as follows:
$b=1/F_c$. \IS{\citet{Repp:2020vr} used a statistical two-state Ising
  bias model, which agrees with this result in the zero temperature
  limit.}  However, it is unclear how accurately the high-density
regions of DM represent real galaxies, particularly for low
luminosities.  The fraction of matter in the clustered and
approximately unclustered populations can be found in the numerical
simulation of the evolution of the cosmic web and determined
independently of bias values from the correlation function. This
factor yields an additional constraint on the bias parameter, besides
the traditional correlation or power spectrum analyses.

The evolution of the biasing properties with cosmic epoch can be
studied using numerical simulations of the cosmic web.  One of the
first studies of the evolution of the two-point correlation function
(CF) of cold dark matter (CDM) universes was by
\citet{Davis:1985}. The authors found that the amplitude of CF
increases with time.  Authors assumed that galaxies form in
high-density peaks of the underlying matter density field and that the
correlation length $r_0$ of simulated galaxies exceeds the correlation
length of matter by a factor of 2.4 at expansion parameter $a=1.4$. A
similar biasing recipe was used by \citet{Einasto:1987kw} and
\citet{Gramann:1987ai,Gramann:1988} in CDM and $\Lambda$CDM
universes. Using $\Lambda$CDM simulations and the $b=1/F_c$ constraint
\citet{Einasto:1999ku} obtained for the bias parameter of galaxies
$b_{c}=1/F_{c}=1.32 \pm 0.13$.  \citet{Cen:1992kx},
\citet{Blanton:1999aa} and \citet{Cen:2000} compared distributions of
galaxies and matter using hydrodynamical simulations of the formation
and evolution of galaxies.  The authors found that there are no
galaxies in regions of low spatial density of matter. The region of
total density, including galaxies, starts at a mean density
$\rho \approx 3$.  \citet{Cen:2000} found for the bias parameter in
the present epoch a value $b=1.35$.

Using Millennium simulations by
\citet{Springel:2006lp}. \citet{Springel:2005aa} found for
$L_\star$-galaxies bias values $b=2.7$ for $z=3$, and $b=0.9$ for
$z=0$, i.e. at the present epoch $L_\star$-galaxies are antibiased.
More recent hydrodynamical simulations of galaxy formation allowed us
to include essential physical processes to simulate and follow
galaxies' formation and evolution in more detail. Examples of such
simulations are Illustris \citep{Vogelsberger:2014aa,
  Vogelsberger:2014ab}, EAGLE \citep{Schaye:2015aa}, HorizonAGN
\citep{Dubois:2016aa}, and The Next Generation (TNG) series of
hydrodynamical simulations by \citet{Springel:2018aa} and
\citet{Pillepich:2018aa, Pillepich:2018ab}.  The first analysis of
matter and galaxy clustering using TNG100 and TNG300 simulations was
made by \citet{Springel:2018aa}.  \citet{Springel:2018aa} calculated
bias parameter values for a range of evolution epochs from $z=5$ to
$z=0$, separately for galaxies of different stellar masses,
star formation rate, and for halos of various masses.  For the present
epoch $z=0$ \citet{Springel:2018aa} found a value $b\approx 1.02$.  As
noted by \citet{Springel:2018aa}, this low value of the bias parameter
is consistent with analytic models by \citet{Mo:1996aa},
\citet{Sheth:1999ab} and \citet{Tinker:2010fk}, and with the observed
CF and bias parameter of SDSS galaxies, as found by \citet{Li:2009aa}.

This short overview of previous determinations of the bias parameter
shows that bias values for $L_\star$ type galaxies are concentrated
around two values, $b_\star\approx 1$ and $b_\star\approx 1.3$.  The
smaller value is obtained from galaxy clustering studies, either
observational or using modern hydrodynamical simulation of the
evolution of galaxies.  The higher value is found in DM-only
numerical simulations and early hydrodynamical simulations. The
difference between these values is much more significant than possible
random errors.  However, biasing is not a phenomenon which can be
characterised by CFs or power spectra alone.  It is closely related to
the division of matter into clustered and approximately unclustered
populations of dark and baryonic matter. Thus, the bias parameter
depends on the luminosity of galaxies and the fraction of matter in
the clustered and approximately unclustered populations. The first
factor is well-known from the first studies of the bias phenomenon by
\citet{Kaiser:1984} and \citet{Bardeen:1986}. The second factor is
related to the evolution of the cosmic web.

In this paper, our main goal is to characterise the differences in the
bias evolution of galaxies and DM. To do this, we determine the bias
parameters of galaxies and clustered DM particles as test objects in
the correlation analysis. We use the Illustris TNG
simulations of the evolution of the cosmic web, where data on
simulated galaxies and DM particles are available.  The TNG
series uses three box sizes, $L_0 = 35,~75,~205~\Mpc$, called TNG50,
TNG100 and TNG300.  The latter two are large enough to study
the evolution of galaxies in the cosmic web.  We shall use TNG100 and
TNG300 simulations within cosmic epochs corresponding to redshifts
from $z=5$ to the present epoch $z=0$.  These simulations have sufficient
mass resolutions to track the evolution of galaxies
substantially below $L_\star$.  We analyse the clustering properties
of simulated galaxies of various luminosities and DM particle
samples with varying particle density limits.  In addition to the
correlation analysis, we study the evacuation of DM particles from
voids, which yields independent information on the evolution of the
bias parameter.

As an independent check, we also analyse the evolution of the bias
parameter using Horizon Run 5 (HR5) simulations by \citet{Lee:2021aa} and \citet{Park:2022aa}.

The paper is structured as follows. In Section 2 we give an overview
of TNG and HR5 galaxy and DM samples and methods to calculate
correlation functions and bias properties.  In Section 3 we analyse
the evolution of the physical properties of simulated galaxies.  In
Section 4 we describe the correlation and bias functions of simulated
galaxies and DM density-limited samples. We analyse the evolution of
clustering properties of simulated galaxies of various luminosity and
DM density-limited samples.  In Section 5 we discuss our results and
compare them with earlier studies. Section 6 summarises the
conclusions of our study.

\section{Data and methods}

In this Section, we describe TNG100,  TNG300 and HR5 simulations used in our
analysis, and methods to calculate correlation and bias functions of
DM and simulated galaxy samples.

\subsection{TNG simulations}

Basic data on TNG100 and TNG300 simulations are given in
\citet{Springel:2018aa} and \citet{Pillepich:2018aa}. High-resolution
versions of these simulations, TNG100-1 and TNG300-1, have particle
numbers $1\,820^3$ and $2\,500^3$, respectively, both for DM and gas
particles, and the same number of cells. Low resolutions versions,
TNG100-3 and TNG300-3, have $455^3$ and $625^3$ cells and numbers of
DM and gas particles, respective masses of DM and gas particles are
given in Table~\ref{Tab1}.  TNG simulations are made for cosmology
parameters $\Omega_m= \Omega_{DM} + \Omega_b = 0.3089$,
$\Omega_b=0.0486$, $\Omega_\Lambda = 0.6911$, and Hubble constant
$H_0=100~h\,$km\,s$^{-1}$Mpc$^{-1}$ with $h=0.6774$.  Initial conditions
were generated for the epoch $z=127$ using a linear theory power
spectrum with a normalisation $\sigma_8=0.8159$ and spectral index
$n_s=0.9667$.  For conformity with literature, we express lengths in
units $\Mpc$.

{\scriptsize 
\begin{table}
\caption{Parameters of simulations }
\centering
\begin{tabular}{lccc}
\hline  
  Simulation   & $L_0~[\Mpc]$&$m_{DM}~[h^{-1}M_\odot]$&$m_{gas}
                                                  ~[h^{-1}M_\odot]$\\   
(1)&(2)&(3)&(4)\\
  \hline  \\
  TNG100-1& 75&$7.5\times10^6$&$1.4\times10^6$\\
  TNG100-3& 75&$4.8\times10^8$&\\
 TNG300-1& 205&$4.0\times10^7$&$7.4\times10^6$\\
  TNG300-3& 205&$4.5\times10^9$&$7.0\times10^8$\\
 \label{Tab1}                         
\end{tabular} \\
{Columns give: (1) name of simulation; (2) box size in
  $\Mpc$; (3) DM particle mass; (4) gas particle mass.
  } 
\end{table} 
}

{\scriptsize 
\begin{table}
\caption{Number  of DM particles and galaxies in simulations }
\centering
\begin{tabular}{lcrr}
\hline  
  Simulation   & $z$& $N_{DM}$&$N_{gal}$\\   
(1)&(2)&(3)&(4)\\
  \hline  \\
 TNG100-1& 0&$455^3$&337\,261\\
 TNG100-1& 0.5&$455^3$&410\,155\\
 TNG100-1& 1&$455^3$&492\,655\\
 TNG100-1& 2&$455^3$&680\,244\\
 TNG100-1& 3&$455^3$&836\,537\\
 TNG100-1& 5&$455^3$&762\,174\\
 TNG100-1& 10&$455^3$&112\,084\\
  \\
  TNG300-1& 0&$625^3$&1\,947\,928\\
  TNG300-1& 0.5&$625^3$&2\,324\,925\\
  TNG300-1& 1&$625^3$&2\,678\,779\\
  TNG300-1& 2&$625^3$&3\,183\,544\\
  TNG300-1& 3&$625^3$&3\,218\,212\\
  TNG300-1& 5&$625^3$&2\,043\,334\\
  TNG300-1& 10&$625^3$&65\,012\\
\label{Tab2}                         
\end{tabular} \\
{Columns give: (1) name of simulation; (2) simulation epoch
  $z$; (3) number of DM particles;
  and (4) number of galaxies.
  } 
\end{table} 
}

We downloaded from the TNG site subhalo data for simulations TNG100-1,
TNG300-1 and TNG300-3, and DM data for TNG100-3 and TNG300-3. Files
for DM data for TNG100-1 and TNG300-1 are too large for
downloading. Experience with Uchuu simulation shows that randomly
selected DM particle samples yield correlation analysis results
fully compatible with DM particle samples
\citep{Ishiyama:2021te}.

\begin{figure*}
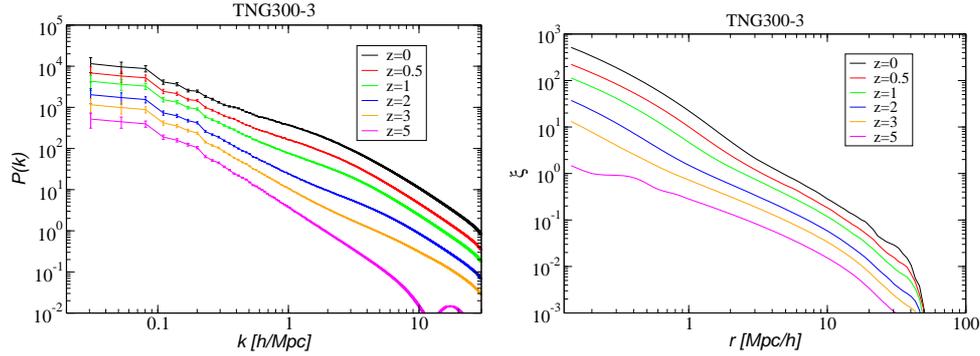

\centering 
\hspace{2mm}
\resizebox{0.35\textwidth}{!}{\includegraphics*{TNG_specter.eps}}
\hspace{2mm}
\resizebox{0.35\textwidth}{!}{\includegraphics*{TNG_CF_DM_allz.eps}}
\caption{Power spectra and CFs of DM in TNG300-3
  simulations for redshifts $z=0,~0.5,~1,~2,~3,~5$ are shown in left
  and right panels, respectively. }
\label{fig:Fig1} 
\end{figure*} 

For subhalos, we extracted $x,~y,~z$-coordinates,
subhalo total and stellar masses, and luminosities of simulated
galaxies in $g$ and $r$ colours from the TNG website.  Our primary analysis is done in the
$r$-band, but we also analysed $g-r$ colours and masses of subhalos.
Galaxy absolute magnitudes in $r$-band were used as labels to select simulated 
galaxy samples for further analysis. The derivation of galaxy magnitudes is described by \citet{Pillepich:2018aa}. 
Numbers of DM particles and simulated galaxies for all redshifts used
in our analysis are given in Table~\ref{Tab2}.  The number of subhalos
for simulation TNG300-3 is much larger than the number of galaxies,
i.e. there exists a large number of subhalos with no galaxy data.  
This simulation's resolution is insufficient to find galaxy data
for low-mass subhalos.  For this reason, galaxies of the simulation
TNG300-3 were not used in the final analysis.  For simulations
TNG100-1 and TNG300-1, almost all subhalos contain galaxies. Thus we
give for these simulations in Table \ref{Tab2} only numbers of
simulated galaxies.

For simulations TNG100-3 and TNG300-3, we extracted DM particle
$x,~y,~z$-coordinates and local densities. The local total comoving
mass density is estimated using the standard cubic-spline SPH kernel
over all particles/cells within a comoving radius of the sphere
centred on this particle enclosing the $64\pm 1$ nearest DM
particles \citep[for details see][]{Nelson:2015aa,Springel:2018aa}. Densities are given in Solar masses per cubic comoving kpc, corresponding to low numerical values for the 
density. We used these densities as labels to select
DM particles of the clustered population, $\rho \ge\rho_0$, where the
limiting density was selected to obtain clustered DM particle samples
for various density thresholds. This method to select particles was
applied earlier by \citet{Jensen:1986aa}, \citet{Einasto:1991fq},
\citet{Szapudi:1993aa}, and \citet{Little:1994rt}. This model to
select particles is similar to the Ising model, discussed by
\citet{Repp:2019ti, Repp:2019wl}. Also, this method allows for finding the
fraction of DM particles in the clustered and unclustered population.
  
\subsection{HR5 simulations}

Hydrodynamical Horizon Run 5 (HR5) simulations are described by
\citet{Lee:2021aa} and \citet{Park:2022aa}.  Simulations were run with
cosmological parameters: $\Omega_m = 0.3$, $\Omega_b=0.047$,
$\Omega_\Lambda = 0.7$, and Hubble constant $h_0=0.684$; in a cubic
box of comoving physical length $L=1049$~cMpc.  Within this cube, a
high-resolution cuboid with a volume $1049\times119\times127$~cMpc$^3$
was selected, which allowed resolving kiloparsec physical
scales.  Simulations started at redshift $z=200$ and finished at
redshift $z=0.625$.  The whole box has, at redshift $z=0.625$,
$7.7\times 10^9$ DM particles and $2.2\times 10^9$ star particles. In
the high-resolution region at epoch $z=0.625$ there are 290\,086
galaxies of mass $M_\star \ge 10^9~M_\odot$. The volume of the
high-resolution cuboid is $(204~\Mpc)^3$, almost equal to the volume
of the TNG300-3 simulation cube.  The simulation TNG300-1 has 290\,061
galaxies with absolute magnitudes $M_r\le-18.41$.

For the present study, one of the coauthors (JK) calculated galaxy and DM correlation
functions for simulation epochs
$z=0.625,~1,~2,~3,~5,~7$, and for galaxies of stellar masses $M_\star \ge
10^9,~\ge 3\times10^9, \ge 10^{10}~M_\odot$.   CFs were calculated with the
\citet{Landy:1993ve} estimator  in 30 logarithmic bins up to
separation $r=140$~cMpc.

\subsection{Calculation of the correlation and bias functions}

Simulated DM files contain a large number of particles. Thus we
applied the \citet{Szapudi:2005aa} grid-based method with
$N_{\mathrm{grid}}=2048^3$ for finding correlation and bias
functions on sub-megaparsec scales.   CFs for simulation TNG100-1 were found up
to separations $r_{max}=37.5~\Mpc$ with 90 logarithmic bins, and for
simulation TNG300-1 up to separations $r_{max}=100~\Mpc$ with 98
logarithmic bins.  For consistency with the DM samples, we used this method
also for simulated galaxies, applying the same bins as for DM samples.
This allows easy computation of bias functions.

We calculated the CFs $\xi(r)$ for all samples of TNG simulations for
two sets of data.  In the first set, we used as test objects simulated
galaxies. CFs were calculated with a series of galaxy luminosity
limits, $M_r$. In the second set, we used as test objects DM
particles with a series of particle density limits,
$\rho_0$.  We always used galaxy/particle samples in cumulative form,
i.e. all galaxies/particles equal to and above some limit are considered.

The ratio of CFs of the galaxy (particle) samples with luminosity limits
$M_r$ (particle density limits $\rho_0$) to correlations functions of
all  DM,  both at identical
separations $r$,  defines the bias function $b(r,M_r)$:
  \begin{equation}
  b^2(r, M_r) = \xi(r,M_r)/\xi_{DM}(r),
\label{bias}  
\end{equation}
and a similar formula for $b(r,\rho_0)$, where the limiting luminosity
$M_r$ is replaced by the particle density limit $\rho_0$.  Bias
depends on the luminosity $M_r$ of galaxies (particle density limit
$\rho_0$), used in the calculation of CFs.

Bias functions have a plateau at $6 \le r \le 20~\Mpc$, see
Fig.~\ref{fig:Fig5} below.  This feature is similar to the plateau
around $k \approx 0.03$~$h$~Mpc$^{-1}$ of relative power spectra
\citep{Einasto:2019aa}.  Following \citet{Einasto:2020aa,
  Einasto:2021ti, Einasto:2023aa}, we use this plateau to measure the
relative amplitude of the CF, i.e. of the bias
function, as the bias parameter,
\be
b(M_r)= b(r_0,M_r),
\label{bias2}
\ee
and a similar formula, where $M_r$ is replaced by particle density
limit $\rho_0$, and $r_0$ is the value of the separation $r$ to
measure the amplitude of the bias function.  We calculated for all
samples bias parameters for the comoving separation,
$r_0=r_{10}=10~\Mpc$, as functions of the galaxy absolute magnitude in
$r$ colour, $M_r$, or particle density limit $\rho_0$ for DM
simulations.  At smaller distances, bias functions are influenced by
the distribution of particles and galaxies in halos, and at larger
distances, the bias functions have wiggles, which makes difficult the
comparison of samples with various galaxy luminosity limits.

\begin{figure*}
\centering 
\hspace{1mm}
\resizebox{0.98\textwidth}{!}{\includegraphics*{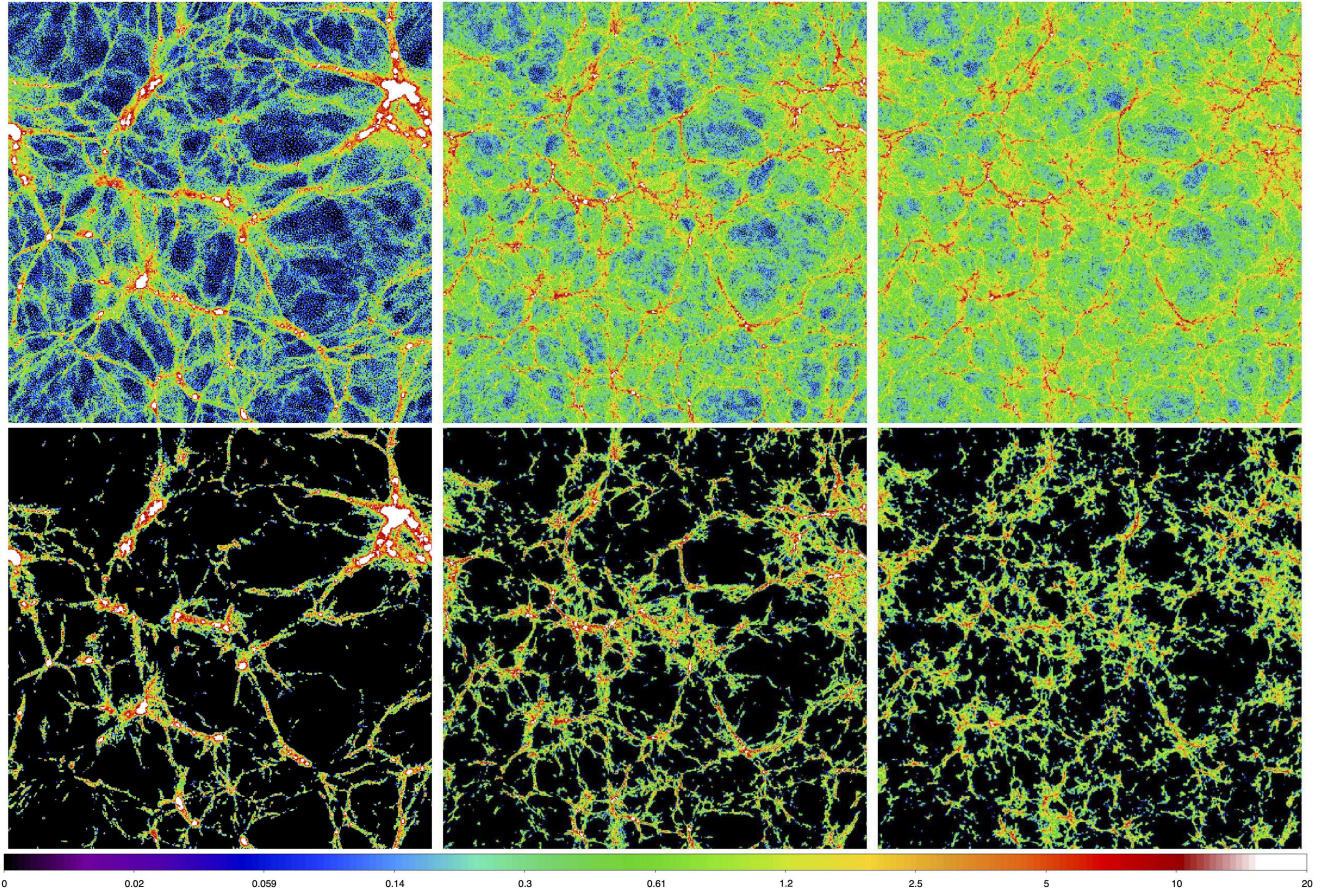}}
\caption{Density fields of DM of the TNG100-3 simulation. The top panels
  show density fields of the full DM, and the bottom panels show density
  fields of the clustered population of DM using DM particles with
  densities $\log\rho\ge\log\rho_0=-7.8$, i.e. all particles of the
  clustered population.  The left, central and right panels show density
  fields at epochs $z=0$, $z=2$ and $z=5$. The densities are given in
  the logarithmic scale to see better the distribution of faint
  filaments.  }
\label{fig:Fig2} 
\end{figure*}

\begin{figure*}
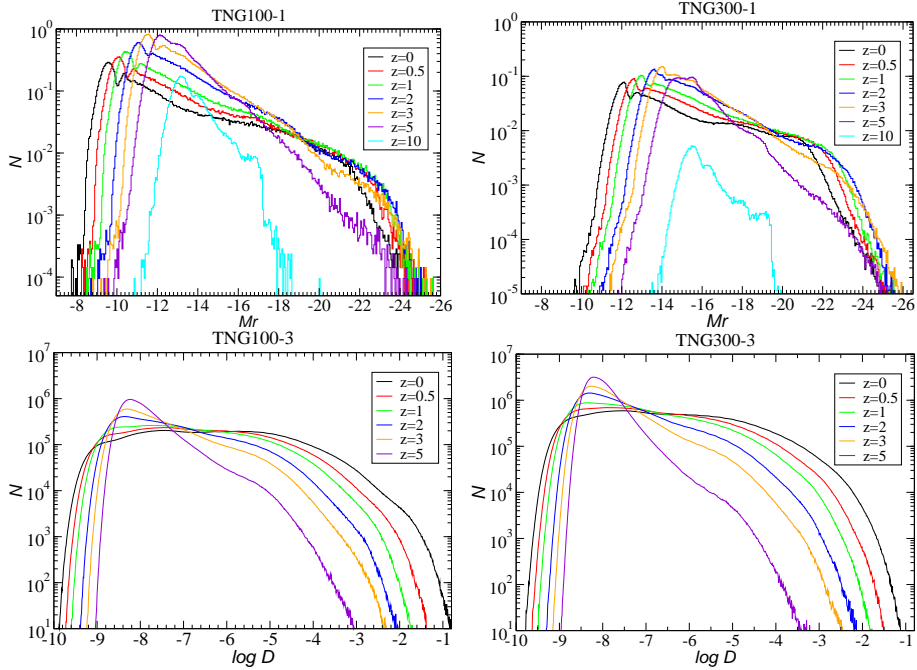

\centering 
\hspace{1mm}
\resizebox{0.33\textwidth}{!}{\includegraphics*{TNG100_Mr-distr.eps}}
\hspace{1mm}
\resizebox{0.33\textwidth}{!}{\includegraphics*{TNG3_1_Mr_diff_05_distr.eps}}\\
\hspace{1mm}  
\resizebox{0.33\textwidth}{!}{\includegraphics*{TNG100-3_DMpart_densdiff4_zall.eps}}
\hspace{1mm}  
\resizebox{0.33\textwidth}{!}{\includegraphics*{TNG300_3_mass_distr.eps}}
\caption{Top panels: Differential galaxy luminosity distribution in
  photometric system $r$. Panels from left to right are for TNG100-1
  and TNG300-1 simulations.  Colour codes show distributions for
  various $z$. The bottom panels show the differential distribution of
  DM particles of TNG100-3 and TNG300-3 as a function of the total
  density at the location of particles, $\log\rho_0$.  }
\label{fig:Fig3} 
\end{figure*}

\section{Evolution of physical properties of simulated galaxies}

In this Section, we describe the evolution of the physical properties of
simulated galaxies and DM samples.  We begin with a description of the
evolution of power spectra and CFs of DM samples,
thereafter we discuss the evolution of spatial distribution and 
luminosity functions of simulated galaxies.

\subsection{Evolution of DM power spectra and CFs}

We calculated the power spectra and CFs of DM particle samples for 
the simulation  TNG300-3. To find power spectra we applied standard
procedure as discussed by \citet{Jing:2005aa}.  For CFs, we used
the \citet{Szapudi:2005aa} method as discussed above.  The evolution
of power spectra and correlation functions of DM is presented in
Fig.~\ref{fig:Fig1}.  As expected, the amplitudes of both functions
increase considerably with time. The growth of amplitudes of both
functions describes the growth of amplitudes of density fluctuations
with time.

Fig. ~\ref{fig:Fig1} shows that CFs of DM deviate considerably from
a simple power law. On large separations, CFs can be approximated by a
power law with exponent $\approx -1.8$.  In this separation range, CFs
describe fractal properties of the distribution of halos, as discussed
by \citet{Einasto:2020aa}.  On smaller separations, CFs describe the
distribution of DM particles in halos, as discussed among others by
\citet{Springel:2018aa} and \citet{Einasto:2020aa,
  Einasto:2023aa}. The characteristic diameters of halos determine the transition between these regimes at a separation $r \approx 3~\Mpc$.

\subsection{Evolution of the spatial distribution of galaxies}

We calculated density fields for DM and galaxies. We calculate the nearest grid point (NGP) 
DM density field from a grid of size $N^3_{\mathrm{grid}}$ 
with the same resolution as the number of particles,
$N_{\mathrm{part}}^3$.  The NGP method was used to find DM density
fields of the TNG100-3 simulation with $N_{\mathrm{grid}} =455$ and
$N_{\mathrm{part}} =455^3$.  Results are presented in
Fig.~\ref{fig:Fig2} for the TNG100-3 simulation in $x,~y-$coordinates
in a sheet of a thickness of 11 simulation cells, $1.8~\Mpc$, across a
massive cluster.  The top panels are for full DM samples, bottom panels
for DM particles with densities $\log\rho\ge\log\rho_0=-7.8$, i.e. all
particles of the clustered population.  Colour codes are for grid
cells of different spatial densities.

The Figure demonstrates the presence of the cosmic web from
the early epoch $z=5$ to the present epoch $z=0$.  All
principal elements of the cosmic web are present already at the epoch
$z=5$, see also \cite{Springel:2006lp, Springel:2018aa},
\cite{Park:2022aa} and \citet{Asgari:2023aa}. This result is consistent with Table~\ref{Tab2}
showing that the number of
subhalos/galaxies did not change considerably during this period.  The number of resolved galaxies, including dwarf satellite galaxies, reaches a maximum at $z=3$, and decreases slightly
thereafter. The contraction of
superclusters -- low-mass systems flow towards central clusters of
superclusters is also visible in these plots. A similar effect was noted by \citet{Einasto:2019ac,
  Einasto:2021wa}.

The top panels of Fig.~\ref{fig:Fig2} show the presence of faint DM
filaments, absent in the bottom panels in the density fields of the
clustered matter. Here only strong galaxy filaments exist, and most
of the volume has zero density. A similar difference is seen in plots
of DM and galaxy/halo distributions  by \cite{Springel:2006lp,
  Springel:2018aa}, \cite{Park:2022aa} and \citet{Asgari:2023aa}.  Already the visual
inspection of the bottom panels shows the increase of the fraction of the
volume of zero-density cells with time (decreasing $z$).

\subsection{Evolution of luminosity functions  of galaxies}

Differential luminosity functions of galaxies of TNG100-1 and TNG300-1
simulations are shown on the top panels of Fig.~\ref{fig:Fig3}.  This
Figure demonstrates clearly the presence of differences in the number
of galaxies in these simulations.  As expected, the number of galaxies
per unit magnitude interval and cubic megaparsec is the largest in the
TNG100-1 simulation.  The increase in the number of galaxies in the
TNG100-1 simulation comes essentially from dwarf galaxies with
luminosities $M_r \ge -18.0$, where the number of dwarf galaxies per
cubic megaparsec is up to ten times larger than in the simulation
TNG300-1. The faint end tail of luminosity distribution of the
TNG100-1 simulation is two magnitudes fainter than that of the simulation
TNG300-1.

\begin{figure*}
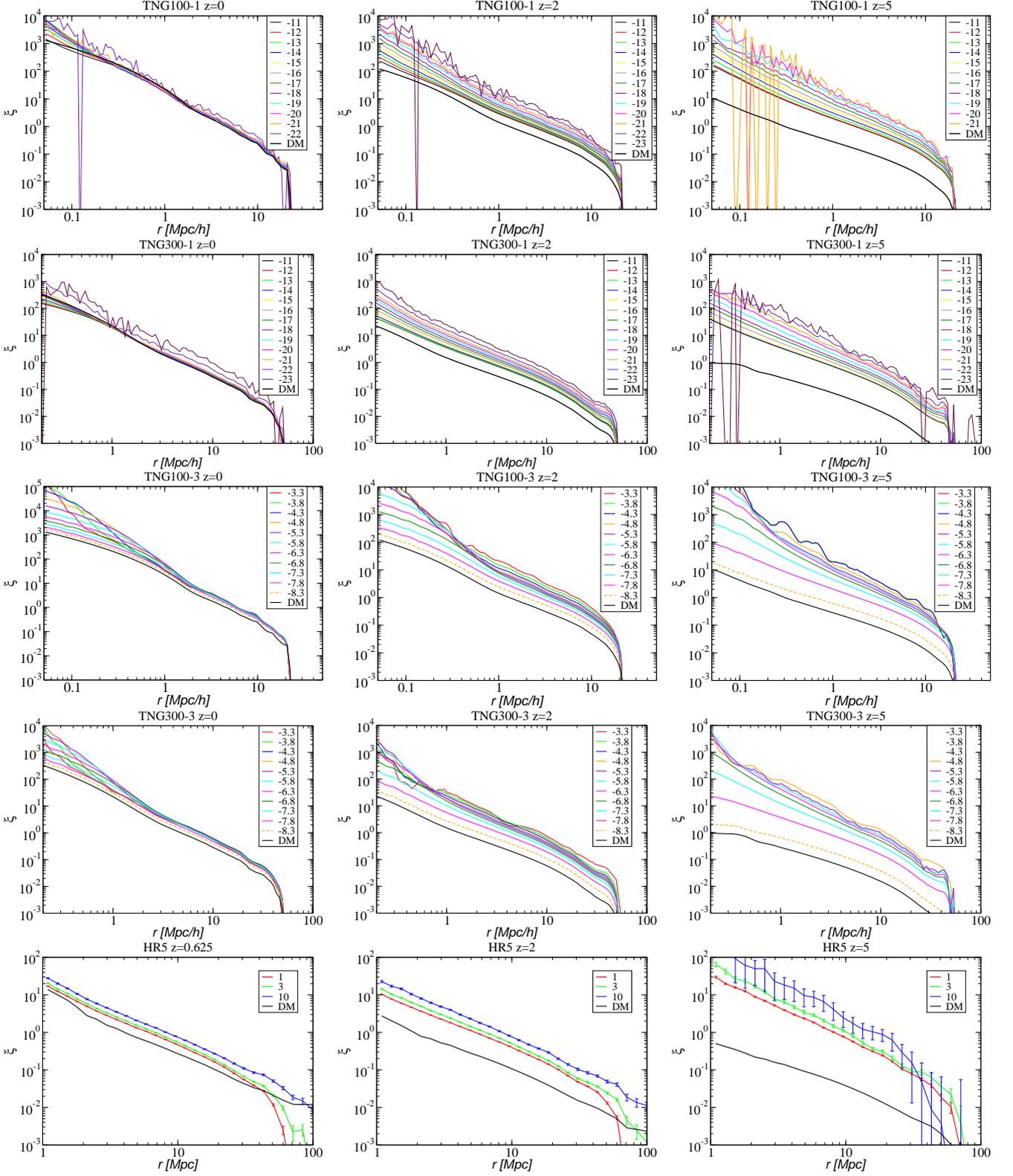

\centering 
\hspace{1mm}  
\resizebox{0.32\textwidth}{!}{\includegraphics*{TNG100-1_CF_2048_z0.eps}}
\hspace{1mm}  
\resizebox{0.32\textwidth}{!}{\includegraphics*{TNG100-1_CF_2048_z2.eps}}
\hspace{1mm}  
\resizebox{0.32\textwidth}{!}{\includegraphics*{TNG100-1_CF_2048_z5.eps}}\\
 \hspace{1mm}  
\resizebox{0.32\textwidth}{!}{\includegraphics*{TNG300-1_CF_2048_z0.eps}}
\hspace{1mm}  
\resizebox{0.32\textwidth}{!}{\includegraphics*{TNG300-1_CF_2048_z2.eps}}
\hspace{1mm}  
\resizebox{0.32\textwidth}{!}{\includegraphics*{TNG300-1_CF_2048_z5.eps}}\\
\hspace{1mm}  
\resizebox{0.32\textwidth}{!}{\includegraphics*{CF_100-3_z0_2048.eps}}
\hspace{1mm}  
\resizebox{0.32\textwidth}{!}{\includegraphics*{CF_100-3_z2_2048.eps}}
\hspace{1mm}  
\resizebox{0.32\textwidth}{!}{\includegraphics*{CF_100-3_z5_2048.eps}}\\
\hspace{1mm}  
\resizebox{0.32\textwidth}{!}{\includegraphics*{CF_300-3_z0_2048.eps}}
\hspace{1mm}  
\resizebox{0.32\textwidth}{!}{\includegraphics*{CF_300-3_z2_2048.eps}}
\hspace{1mm}  
\resizebox{0.32\textwidth}{!}{\includegraphics*{CF_300-3_z5_2048.eps}}\\
\hspace{1mm}  
\resizebox{0.32\textwidth}{!}{\includegraphics*{CF_HR5_z0.625.eps}}
\hspace{1mm}  
\resizebox{0.32\textwidth}{!}{\includegraphics*{CF_HR5_z2.eps}}
\hspace{1mm}  
\resizebox{0.32\textwidth}{!}{\includegraphics*{CF_HR5_z5.eps}}
\caption{CFs of galaxies, $\xi(r)$ for epochs $z=0,~2,~5$, shown in
  the left, central and right panels. The top and second rows are for the
  simulations TNG100-1 and TNG300-1 respectively. Separations $r$ are in comoving units. Magnitude limits in
  $r$ photometric system are shown as symbol labels. The third and fourth
  rows are for TNG100-3 and TNG300-3 DM simulations.  Coloured lines
  show functions for various DM particle density limit $\log\rho_0$.
  The bottom row is for CFs of HR5 simulations for epochs
  $z=0.625,~2,~5$. The stellar mass of galaxies is in units $10^9~M_\odot$.
  Black bold lines show CFs of DM for respective epochs.  }
\label{fig:Fig4} 
\end{figure*} 

Differences between simulations TNG100-1 and TNG300-1 are present 
also in the high-luminosity tail of the distribution at various epochs.  At
the early epoch, $z=10$, the most luminous galaxies have luminosity
$M_r \approx -18.0$ for TNG100-1 and $M_r \approx -20.0$ for TNG300-1.
For both simulations, there is a rapid increase of the luminosity of
the brightest galaxies between redshifts $z=10$ and $z=5$. This increase
is due to the merging of galaxies in the centres of massive halos.  The
difference between simulations TNG100-1 and TNG300-1 
can probably be explained by the larger volume of simulation TNG300-1,
which contains larger modes of density perturbations and allows more
effective merging.  Notice also that the luminosity of most luminous
galaxies with $M_r < -22.0$ increases between redshifts $z=10$ to
$z=3$, and thereafter decreases slightly. The luminosity function
curve for redshift $z=0$ at the high end is lower than for redshifts
$z=2$ and $z=3$.  The most essential difference between simulations at
epochs $z=10$ and $z\le 5$ is in the number of simulated galaxies: at
redshift $z=10$ it is much lower than in smaller redshifts, as seen from
Table ~\ref{Tab2} and Fig. ~\ref{fig:Fig3}.  In further analyses,
we shall use only simulated galaxies at redshifts $z\le 5$.

Bottom panels of Fig.~\ref{fig:Fig3} show the distribution of DM
particle densities at various redshifts $z\le 5$. We see that particle
density distributions at the high density end differ
about three magnitudes  for  redshifts $z=5$ to $z=0$, much more than
for most luminous galaxies $M_r$ at redshifts $z=5$ to $z=0$.

\begin{figure*}
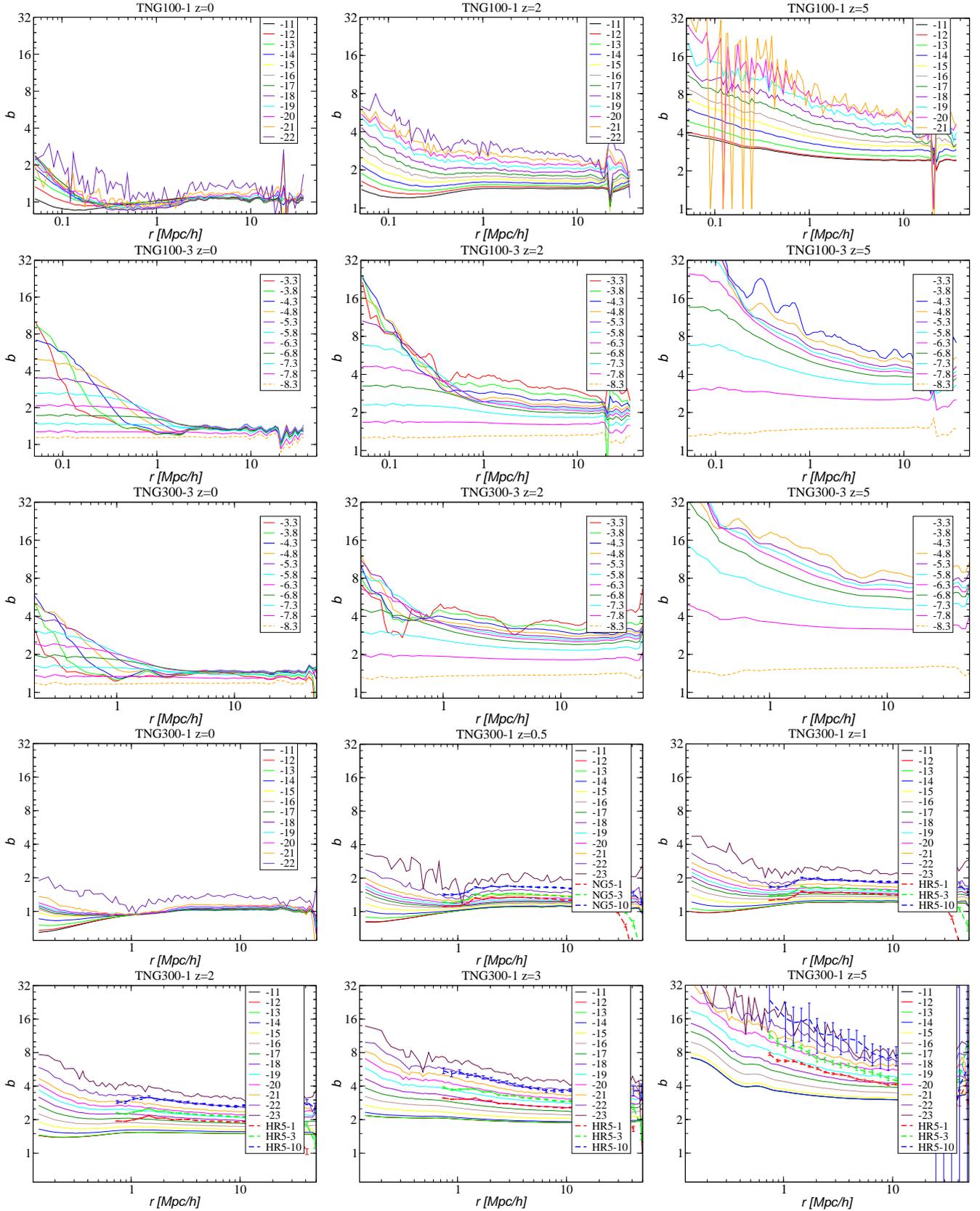

\centering 
\hspace{1mm}  
\resizebox{0.32\textwidth}{!}{\includegraphics*{TNG100-1_bias_2048_z0.eps}}
\hspace{1mm}  
\resizebox{0.32\textwidth}{!}{\includegraphics*{TNG100-1_bias_2048_z2.eps}}
\hspace{1mm}  
\resizebox{0.32\textwidth}{!}{\includegraphics*{TNG100-1_bias_2048_z5.eps}}\\
\hspace{1mm}  
\resizebox{0.32\textwidth}{!}{\includegraphics*{CF_100-3_z0_2048bias.eps}}
\hspace{1mm}  
\resizebox{0.32\textwidth}{!}{\includegraphics*{CF_100-3_z2_2048bias.eps}}
\hspace{1mm}  
\resizebox{0.32\textwidth}{!}{\includegraphics*{CF_100-3_z5_2048bias.eps}}\\
\hspace{1mm}  
\resizebox{0.32\textwidth}{!}{\includegraphics*{CF_300-3_z0_2048bias.eps}}
\hspace{1mm}  
\resizebox{0.32\textwidth}{!}{\includegraphics*{CF_300-3_z2_2048bias.eps}}
\hspace{1mm}  
\resizebox{0.32\textwidth}{!}{\includegraphics*{CF_300-3_z5_2048bias.eps}}\\
\hspace{1mm}  
\resizebox{0.32\textwidth}{!}{\includegraphics*{TNG300-1_bias_2048_z0.eps}}
\hspace{1mm}  
\resizebox{0.32\textwidth}{!}{\includegraphics*{TNG300-1_bias_2048_z0.5.eps}}
\hspace{1mm}  
\resizebox{0.32\textwidth}{!}{\includegraphics*{TNG300-1_bias_2048_z1.eps}}\\
\hspace{1mm}  
\resizebox{0.32\textwidth}{!}{\includegraphics*{TNG300-1_bias_2048_z2.eps}}
\hspace{1mm}  
\resizebox{0.32\textwidth}{!}{\includegraphics*{TNG300-1_bias_2048_z3.eps}}
\hspace{1mm}  
\resizebox{0.32\textwidth}{!}{\includegraphics*{TNG300-1_bias_2048_z5.eps}}\\
\caption{Bias functions of galaxies, $b(r)$, for epochs $z=0,~2,~5$
  shown in the left, central and right panels in the first three rows. The
  top row is for simulations TNG100-1. Separations $r$ are in comoving units.
  Magnitude $M_r$ limits are
  shown as symbol labels.  The second and third 
  rows are for TNG100-3 and TNG300-3 DM simulations. Coloured lines
  show functions for various particle density limits $\log\rho_0$.  The
  limit $\log{\rho_0}=-8.3$ is marked with dashed orange lines, it
  corresponds to DM particles, not associated with galaxies. Two
  bottom rows are for bias functions of TNG300-1 simulations for
  epochs $z=0,~0.5,~1,~2,~3,~5$, superposed with bias functions of HR5
  simulations for similar epochs (epoch $z=0.625$ is shown in panel
  for $z=0.5$). Star masses of galaxies are in units
  $10^9~M_\odot$. Level $b=1$ is the bias function of DM.  }
\label{fig:Fig5} 
\end{figure*}

\section{Evolution of clustering properties of luminosity and
  particle density limited samples of TNG simulations}

The Section begins with a description of the evolution of
correlation and bias functions with time. We show also the evolution
of correlation and bias functions of galaxies of HR5 simulations.
Thereafter we describe the evolution of bias parameters with cosmic
epoch, and its dependence on the luminosities of galaxies. Finally, we
describe the determination of bias parameters of the faintest galaxies and
the fraction of matter in the clustered population.

\subsection{Evolution of  CFs of TNG   simulation galaxies and
  particle density limited DM samples, and CFs of galaxies of HR5 simulations} 

We calculated CFs of galaxies and DM particle density limited samples
for all selected simulations. For all samples, we applied the
\citet{Szapudi:2005aa} method.  As a reference sample, we used DM
particles from simulations TNG100-3 and TNG300-3.  In
Fig.~\ref{fig:Fig4} we show CFs of galaxies for TNG100-1 and TNG300-1
simulations. For all samples a large range of limiting luminosities
$M_r$ was applied from $M_r=-11.0$ to $M_r=-23.0$. The top and second rows
are for the simulations TNG100-1 and TNG300-1, left, middle and right
panels are for evolutionary epochs $z=0$, $z=2$ and $z=5$,
respectively.  Bold black lines show DM CFs, and coloured lines present
galaxy CFs for various $M_r$ limits, shown as labels. 

\begin{figure*}
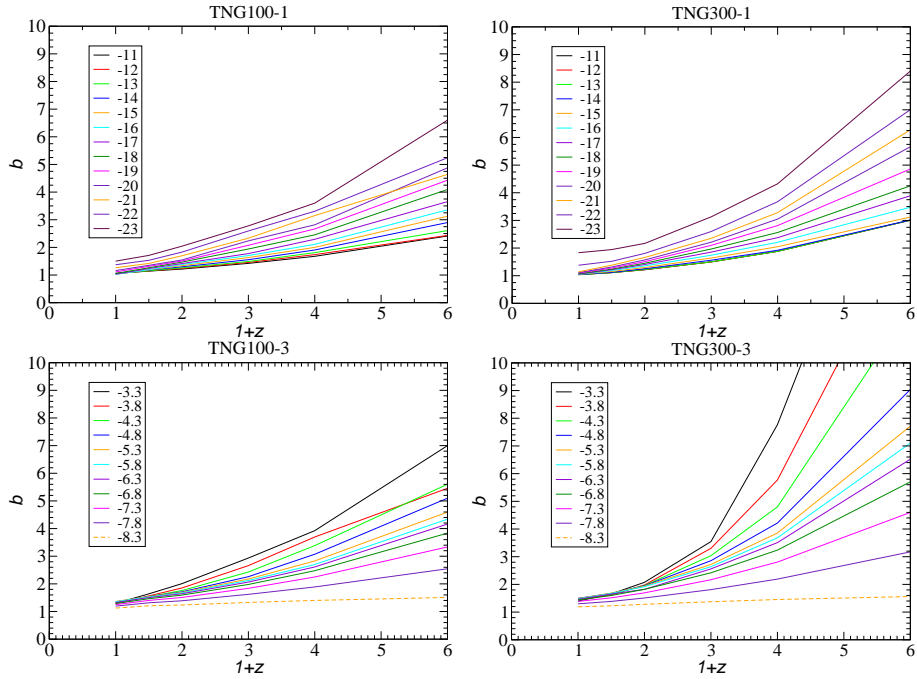

\centering 
   \hspace{1mm}  
\resizebox{0.33\textwidth}{!}{\includegraphics*{TNG100-1_bias_zall.eps}}
\hspace{1mm}  
\resizebox{0.33\textwidth}{!}{\includegraphics*{TNG300-1_bias_zall.eps}}\\
\hspace{1mm}  
\resizebox{0.33\textwidth}{!}{\includegraphics*{TNG100-3_bias10conv_zall.eps}}
\hspace{1mm}  
\resizebox{0.33\textwidth}{!}{\includegraphics*{TNG300-3_bias10conv_zall.eps}}
\caption{Evolution of bias parameter values with the epoch of simulations
  $z$.  The top left and right panels show data for simulations TNG100-1
  and TNG300-1. Magnitude limits in $M_r$ photometric system are shown
  as symbol labels.  The bottom panels show the evolution of the bias parameter of
  DM samples of simulations TNG100-3 and TNG300-3 with various limits
  of substructure density $\log{\rho_0}$.  The limit
  $\log{\rho_0}=-8.3$ is marked with dashed orange lines, it
  corresponds to DM particles, not associated with galaxies. Level
  $b=1$ is the bias of DM.  }
\label{fig:Fig6} 
\end{figure*}

Fig.~\ref{fig:Fig4} shows that CFs of galaxies for $M_r$
limits form sequences of increasing amplitude with increasing
luminosity limits. The number of galaxies in most samples is large,
thus random errors of CFs are very small. Only for the brightest galaxies, errors
are larger and CFs have visible scatter. 
The Figure shows also that for the present epoch $z=0$ CFs
of the faintest galaxies almost coincide with CFs of
DM. For earlier epochs, there exists a gap in the amplitudes of CFs of galaxies
relative to DM, which increases with the simulation epoch  $z$. This is the
biasing effect, discussed in detail in the next subsection.

In the third and fourth rows of Fig. ~\ref{fig:Fig4}, we present CFs of
particle density limited DM samples of TNG100-3 and TNG300-3
simulations for epochs $z=0,~2,~5$.  Here we used particle density
limits $\rho_0$ in $\log{\rho_0}$ units, starting from
$\log{\rho_0}=-8.3$.  The limit $\log{\rho_0}=-8.3$ corresponds to DM
particles, not associated with galaxies at the epoch $z=0$; the
corresponding CFs  are  marked with  dashed orange lines. DM samples,
corresponding to the faintest simulated galaxies at $z=0$, have the limit
$\log{\rho_0}=-7.8$. DM samples corresponding to most luminous
galaxies have at present epoch $\log{\rho_0}\approx -3.3$.

Bottom panels of Fig.~\ref{fig:Fig4} show CFs of galaxies of HR5
simulations for epochs $z=0.625,~2,~5$, calculated with the
\citet{Landy:1993ve} estimator for separations from 1 to 100
in comoving physical megaparsecs.  Here we used simulated
galaxies with a stellar mass lower limits $M_\star =1,~3,~10$ in units
of $10^9~M_\odot$. 

\subsection{Evolution of bias functions of TNG and HR5 simulation galaxies}

CFs of galaxies divided by CFs of DM define bias functions, see Eq. 
(\ref{bias}). They are shown in Fig.~\ref{fig:Fig5}.  
 The top panels are for simulations TNG100-1, and the left, central and right
panels are for epochs $z=0$, $z=2$ and $z=5$. Various colours are for bias functions of
galaxies of different luminosity limits.
In the second and third  rows of Fig. ~\ref{fig:Fig5}, we present bias functions of particle
density limited DM samples of TNG100-3 and TNG300-3 simulations for
epochs, $z=0,~2,~5$.  As for CFs we used particle density limits
$\rho_0$ in $\log{\rho_0}$ units, starting from $\log{\rho_0}=-8.3$.

Two bottom rows of Fig.~\ref{fig:Fig5} are for bias functions of
TNG300-1 simulations for epochs $z=0,~0.5,~1,~2,~3,~5$, superposed
with bias functions of HR5 simulations for similar epochs, epoch
$z=0.625$ is shown in panel for $z=0.5$. Separations $r$ of HR5
simulations were reduced to units $\Mpc$, using the adopted Hubble
constant $h_0=0.684$.  Star masses of galaxies are in units
$10^9~M_\odot$.

Bias functions, presented in Fig.~\ref{fig:Fig5}, have three important
properties.  The first property is: bias function curves for galaxies
of low luminosity, $M_r \ge -18.0$, are almost identical. We discuss
this effect in more detail below.

The second important feature is the shape of bias functions for 
separations, $r \le 5~\Mpc$.  In this separation range bias functions
have larger values than on medium and larger separations.  This is due
to the effect of halos, which have characteristic diameters up to 
$r \approx 5~\Mpc$.  

The third feature is the amplitude at very small separations,
$r \le 1~\Mpc$, for galaxy samples of the lowest luminosities.  Here bias
functions are lower than at higher separations.  This
effect is observed for epochs $z \le 3$.  For the present epoch $z=0$
this means anti-biasing, since $b<1$.  Bias functions of particle
density limited DM samples of simulations TNG100-3 and TNG300-3 do not
have this feature, in this separation interval and particle density
limits $\log\rho_0 \le -6.8$ their bias functions are almost parallel
lines with amplitudes  increasing with increasing limit $\log\rho_0$.

\subsection{Evolution of bias parameters of luminosity and
  particle density limited samples of TNG simulations}

Following \citet{Einasto:2023aa} we define bias parameters as values
of the bias function at separation $r_0=10~\Mpc$. Top panels of Fig.~\ref{fig:Fig6}
present bias parameters for simulations TNG100-1 and TNG300-1 as
functions of redshift $z$.  Different colours show bias parameters for
galaxies of various $M_r$ luminosity.  As we see, bias parameters of
galaxies form smooth curves, $b(M_r)$,  with amplitudes, increasing with
luminosity $M_r$.  With decreasing luminosity bias parameters $b(M_r)$
approach to asymptotic low limits $b(M_r) \rightarrow b_0$.
Low-luminosity limits $b_0$ of simulation TNG100-1 and 
TNG300-1 almost coincide.

\begin{figure*}
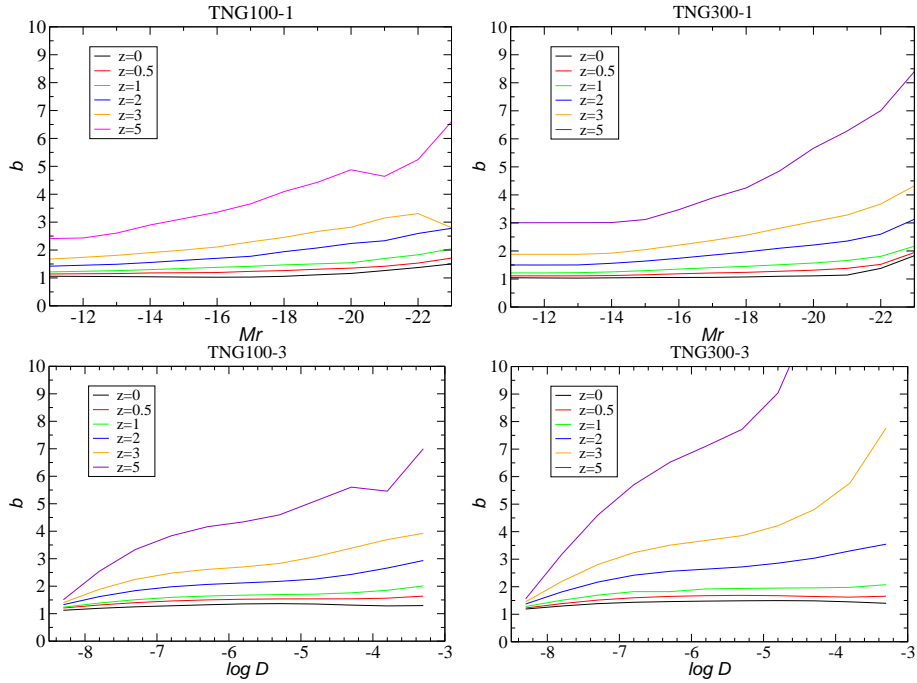

\centering 
\hspace{1mm}  
\resizebox{0.33\textwidth}{!}{\includegraphics*{TNG100-1_bias-Mr_xall.eps}}
\hspace{1mm}  
\resizebox{0.33\textwidth}{!}{\includegraphics*{TNG300-1_bias-Mr_zall.eps}}\\
\hspace{1mm}  
\resizebox{0.33\textwidth}{!}{\includegraphics*{TNG100-3_bias-logD_zall.eps}}
\hspace{1mm}  
\resizebox{0.33\textwidth}{!}{\includegraphics*{TNG300-3_bias-logD_zall.eps}}
\caption{The top left and right panels show the evolution of the bias parameter
  $b$ of simulations TNG100-1 and TNG300-1 as a function of the magnitude limits $M_r$ for various simulation epochs $z$.
  The bottom panels show the evolution of bias parameter $b$ of simulations
  TNG100-3 and TNG300-3 as a function of the logarithm of particle
  density limit $\rho_0$.
  Level $b=1$ is for  DM.  }
\label{fig:Fig7} 
\end{figure*}

\begin{figure*}
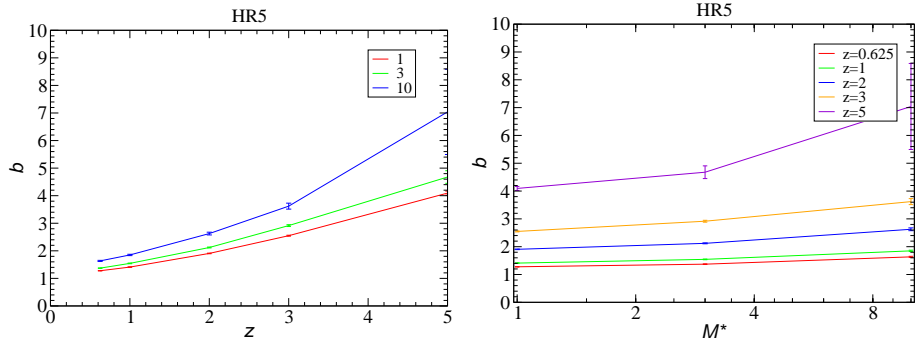

\centering 
\hspace{1mm}  
\resizebox{0.33\textwidth}{!}{\includegraphics*{CF_NG5bias_allz.eps}}
\hspace{1mm}  
\resizebox{0.33\textwidth}{!}{\includegraphics*{CF_NG5bias_allz_rev.eps}}
\caption{Evolution of bias parameter $b(z,M_\star)$ of HR5 simulations. 
   In the left panel, as a function of the epoch $z$, in the right panel, of
  the galaxy stellar mass $M_\star$ in units of $10^9~M_\odot$. 
  }
\label{fig:Fig8} 
\end{figure*}

\begin{figure*}
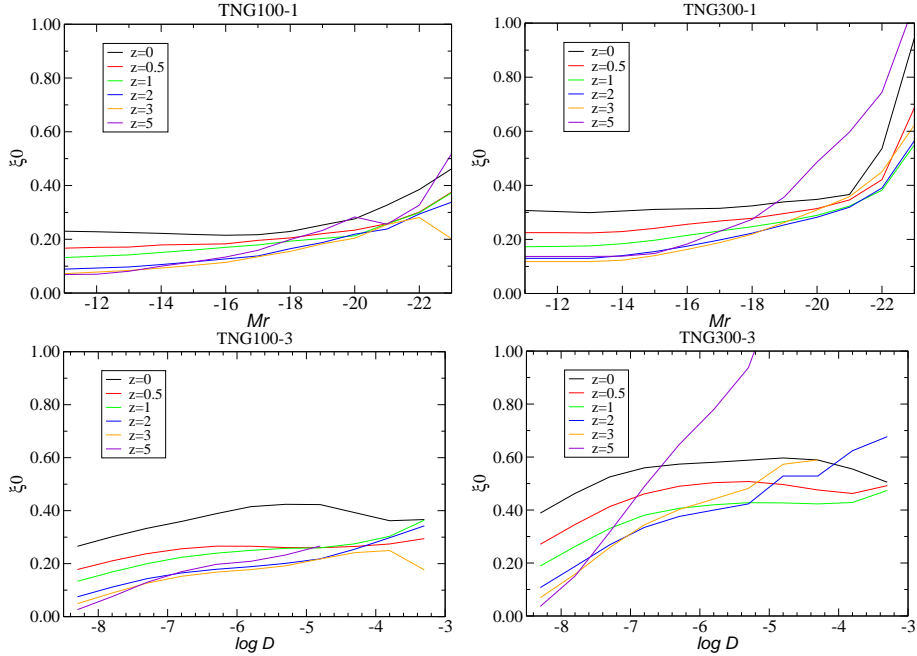

\centering 
\hspace{1mm}  
\resizebox{0.33\textwidth}{!}{\includegraphics*{TNG100-1_Mr-xi0_allz.eps}}
\hspace{1mm}  
\resizebox{0.33\textwidth}{!}{\includegraphics*{TNG300-1_Mr-xi0_allz.eps}}\\
\hspace{1mm}  
\resizebox{0.33\textwidth}{!}{\includegraphics*{TNG100-3_xi10-logD_zall.eps}}
\hspace{1mm}  
\resizebox{0.33\textwidth}{!}{\includegraphics*{TNG300-3_xi10-logD_zall.eps}}
\caption{Top panels plot the evolution of the CF amplitude at fixed
  separation, $\xi(10)$, as a function of magnitude limits $M_r$ for
  simulations TNG100-1 and TNG300-1.  Bottom panels show the evolution
  of CF amplitude $\xi(10)$ of the DM simulations TNG100-3 and
  TNG300-3 for various particle density limits $\log\rho_o$.  }
\label{fig:Fig9} 
\end{figure*}

The bottom panels of Fig.~\ref{fig:Fig6} show the evolution of bias parameters
of particle density limited DM samples of simulations TNG100-3 and
TNG300-3 for various limits $\log\rho_0$.  Samples with DM particle
density limit $\log\rho_0=-7.8$ correspond to the faintest clustered
population, similar to the faintest galaxies of simulations TNG100-1 and
TNG300-1. The dashed orange line corresponds to DM samples with a limit
$\log\rho_0=-8.3$, i.e. to DM particles below the density limit,
needed to form stars and galaxies.  The Figure shows that in
simulation TNG300-3, the curves for the highest particle density limits are
higher than in simulation TNG100-3.  This difference is due to various
shapes of density  distributions, as seen in the bottom panels of  Fig.~\ref{fig:Fig3}.

\subsection{Bias parameters as functions of luminosities  of
  galaxies and particle density limits}

Top panels of Fig.~\ref{fig:Fig7} present bias parameters $b(M_r)$ as
functions of the luminosity $M_r$ for epochs $z=0$ to $z=5$. As noticed in the previous subsection,
an essential property of bias parameters is that at low and medium
luminosities $b(M_r)$ approaches asymptotically a low-luminosity limit, $b_0$, and rises at
luminosities from $M_r \le -15$ to $M_r \le -18$, corresponding
stellar masses are $1.7\times 10^8\,M_\odot$ and $2.7\times 10^9\,M_\odot$; details depend on
simulation epoch $z$.

The bottom panels of Fig.~\ref{fig:Fig7} show the dependence of bias
parameter on particle density limit $\log\rho_0$ of clustered DM
samples of simulations TNG100-3 and TNG300-3. The
dependence of the bias parameter of clustered DM samples of simulations
TNG100-3 and TNG300-3 is different from the dependence of bias
parameter on luminosities of simulations TNG100-1 and TNG300-1. In the 
clustered DM samples there is no flat region of the bias
function $b(\log\rho_0)$ at low particle density limits, $\rho_0$. Rather,  
the amplitudes of $b(\log\rho_0)$ curves rise continuously with increasing
$\log\rho_0$.  Above a very low particle density
$\log\rho_0 \approx -10$ limit essentially all particles are included, see Fig.~\ref{fig:Fig3}. Thus
the bias parameter should be $b=1$ by definition.  The $b(\log\rho)$ curves really
converge to 1 at $\log\rho_0 \approx -9$.

At a particle density limit $\log\rho_0=-7.8$, the bias parameter values of
DM simulations TNG100-3 and TNG300-3 are almost equal to the bias values
of the faintest galaxies of TNG100-1 and TNG300-1 simulations, c.f., in
the top panels of Fig.~\ref{fig:Fig7}.  At higher particle density limits,
the $b(\log\rho_0)$ curves of the DM selected samples from TNG100-3 and
TNG300-3 are rather similar to the $b(M_r)$ curves of the TNG100-1 and
TNG300-1 simulations.  The basic difference lies in the bias values for
earlier epochs -- here $b(\log\rho_0)$ curves lie higher than $b(M_r)$
curves.  This difference is due to the fact that at earlier epochs
identical particle density limits $\log\rho_0$ correspond to more
luminous galaxies, see Fig.~\ref{fig:Fig3}.

\subsection{Evolution of bias parameters of HR5 simulations}

We show in Fig.~\ref{fig:Fig8} the evolution of the bias parameter
$b(z,M_\star)$ of HR5 simulations.  In the left panel, the bias
parameter is presented as a function of the epoch $z$, in the right
panel, as a function of the mass of simulated galaxies $M_\star$ in units
$10^9M_\odot$.  The evolution of the bias parameter of
HR5 galaxies is close to that of the TNG300-1 simulations presented in previous
Figures.  The main difference is the absence of data for epoch $z=0$
and for very low-mass galaxies.

\begin{figure*}
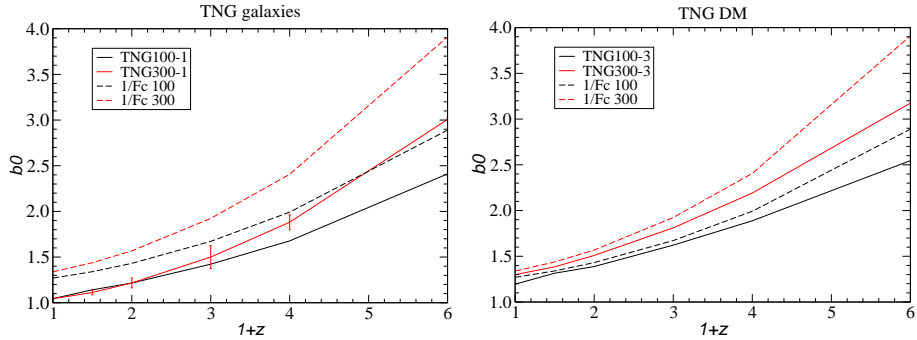

\centering 
\hspace{1mm}  
\resizebox{0.33\textwidth}{!}{\includegraphics*{TNG_zall_bias0.eps}}
\hspace{1mm}  
\resizebox{0.33\textwidth}{!}{\includegraphics*{TNG_DM_bias10conv_zall.eps}}
\caption{Left: Evolution of the bias parameter $b_0$ of the faintest galaxies from simulations TNG100-1 and TNG300-1.  Dashed curves show the
  evolution of the inverse of the fraction of matter in the clustered
  population $b_0=1/F_c$ for simulations TNG100-3 and TNG300-3.
  Right: Evolution of the bias parameter $b_0$ of the faintest galaxies from simulations TNG100-3 and TNG300-3 at a particle density limit
  $\log\rho_0=-7.8$. The dashed curves show the evolution of the inverse
  of the fraction of matter in the clustered population $b_0=1/F_c$
  for simulations TNG100-3 and TNG300-3. }
\label{fig:Fig10} 
\end{figure*}

\subsection{Amplitudes of CFs at fixed separations}

The stable behaviour of the bias parameter of faintest galaxies raises
the question: how amplitudes of CFs at a fixed separation
$\xi_0=\xi(r_0)$ evolve with time.  As discussed above, we use CF
amplitudes at the separation $r_0=10~\Mpc$ to define bias parameters.
We calculated CF amplitudes at separation $r_0=10~\Mpc$ for TNG100-1
and TNG300-1 simulations as functions of the limiting magnitude $M_r$,
and for TNG100-3 and TNG300-3 simulations as functions of limiting
particle density $\log\rho_0$.  These dependencies are shown in the top panels of
Fig.~\ref{fig:Fig9} for  TNG100-1 and TNG300-1 galaxy simulations, and in
bottom panels of Fig.~\ref{fig:Fig9} for  TNG100-3 and TNG300-3 DM simulations. 

Fig. ~\ref{fig:Fig9} shows several important properties of amplitudes
of CFs.  First, the growth of amplitudes of CFs of galaxies with
cosmic epoch $z$ is very modest, much smaller than the growth of
amplitudes of CFs of DM, presented in Fig.~\ref{fig:Fig1}.  This
property is well-known, see \citet{Springel:2006lp, Springel:2018aa}.
Second, we see a big difference between the shape of $\xi_0(M_r)$ and
$\xi_0(\log\rho_o)$ functions of galaxy and DM simulations.  In DM
simulations TNG100-3 and TNG300-3, there exists no  asymptotic  flat region of
$\xi_0(\log\rho_0)$ curves on low particle density limit, present in
low-luminosity regions of $\xi_0(M_r)$ functions of TNG100-1 and
TNG300-1 simulations.  Second, upper limits of DM densities at high
redshifts are much lower than at low redshifts, see
Fig.~\ref{fig:Fig3}, thus $\xi_0(\log\rho_0)$ curves for high
redshifts rise rapidly with increasing $\log\rho_0$, since they
correspond to particles with higher density limits.  The third
essential difference between amplitudes of CFs of galaxy and DM
simulations is the level of $\xi_0$ at low-luminosity regions: in DM
simulations TNG100-3 and TNG300-3, it is higher than in the galaxy
simulations TNG100-1 and TNG300-1.

We also calculated the evolution of correlation lengths $r_0$ of
TNG100-1 and TNG300-1 simulations, $r_0(z)$,  for $M_r$ limited
samples. Results are similar to functions $r_0(M_\star)$, found
by \citet{Springel:2018aa}, and are not presented here.

\subsection{Bias parameter of faintest  galaxies and the fraction of
  matter in the clustered population} 

Our analysis shows that the bias parameter of faintest galaxies,
$b_0$, is a well-defined quantity, almost independent of the
luminosity of galaxies but dependent on the evolutionary epoch $z$.
We determined the asymptotic bias parameter of faintest galaxies,
$b_0$, for TNG100-1 and TNG300-1 simulations.  Bias parameter curves
for luminosities $-11.0 \le M_r \le -14.0$ of simulation TNG100-1 vary
slightly. We accepted the asymptotic value $b_0=1.045$.
Fig.~\ref{fig:Fig10} shows the dependence of $b_0$ on cosmic epoch
$z$.  Our data show that simulations TNG100-1 and TNG300-1 yield similar results for
the asymptotic bias parameter of faintest galaxies $b_0$ similar
results for the present epoch $z=0$. 

The right panel of Fig.~\ref{fig:Fig10} shows the evolution of the
bias parameter $b_0$ of DM simulations TNG100-3 and TNG300-3.  The
particle density $\rho$, extracted from the TNG website, includes all
matter: DM plus baryonic gas and stellar matter.  The $\rho$ value 
for stellar matter is not  given on the TNG website.   We used a two-step procedure to determine
the lower $\rho_0$ limit for stellar matter. First, we found the distribution of the
fraction of matter in the clustered population, $F_c(\rho_0)$, of the
DM-only simulations TNG100-3 and TNG300-3, using the cumulative
distribution of particle densities, see for reference in
Fig.~\ref{fig:Fig3} differential distributions of particle densities,
$N(\log\rho)$.  In the next step, we compared cumulative distributions
of particle densities $F_c(\rho_0)$ with bias parameter distributions
$b(\log\rho_0)$ for various particle density limit $\rho_0$, shown in
Fig.~\ref{fig:Fig7}.  This comparison showed that for particle density
limit $\log\rho_0=-7.8$ functions $b(z)$ and $b_c(z)=1/F_c(z)$ are
very close, and also close to the $b_0$ value for galaxy simulations.
We use this particle density limit $\log\rho_0=-7.8$ as 
the limit for faintest stellar systems  in TNG100-3 and TNG300-3 simulations.
Dashed curves in Fig.~\ref{fig:Fig10} show the inverse of the fraction
of the clustered population, $b_c(z)=1/F_c(z)$ for TNG100-3 and
TNG300-3 simulations.  The Figure shows that at the present epoch $z=0$
both DM simulations yield for $b_0$ very close values, almost identical
to the expected value from the fraction of the clustered population,
$b_0=1/F_c$.

We use the same $b_c(z)=1/F_c(z)$ curves, found for the TNG100-3 and
TNG300-3 DM particle simulations, also for TNG100-1 and TNG300-1
galaxy simulations, shown in the left panel of Fig.~\ref{fig:Fig10} by
dashed curves.  We see that $b_c(z)=1/F_c(z)$ curves of simulations
TNG100-3 and TNG300-3 lie higher than the actual $b_0(z)$ curves of
TNG100-1 and TNG300-1 simulations.

Fig.~\ref{fig:Fig10} shows that the $b_0(z)$
curves of simulations TNG300 for both data types lie higher than those of simulation
TNG100; the difference increases with epoch $z$.  The difference is probably due to varying evolutionary histories of dwarf halos since in the higher resolution simulation TNG100, the number
of dwarf galaxies is much higher than in TNG300, see Fig.~\ref{fig:Fig3}.

\section{Discussion}

In this Section we discuss how bias functions $b(r|z,M_r)$ and bias
parameters $b(z,M_r)$ represent properties of the cosmic web.

\subsection{The shape of evolving correlation and bias functions}

\subsubsection{The shape of correlation and bias functions at large separations}

Correlation and bias functions of simulated galaxies are presented in
Figs.~\ref{fig:Fig4} to \ref{fig:Fig8}.  These Figures show that the bias
parameters $b(z,M_r)$  decrease during the evolution.
This result is not new; it confirms earlier studies by
\citet{Tegmark:1998yq}, \citet{Springel:2005aa, Springel:2018aa},
\citet{Park:2022aa} and \citet{Einasto:2023aa}.
The evolution of the correlation and bias functions of
galaxies in simulations TNG300-1 and HR5 are very similar  as demonstrated in two bottom panels of
Fig.~\ref{fig:Fig5}, where we show bias functions for the whole set of
simulation epochs $z=0$ to $z=5$.  The bias functions of HR5 simulations
lie very close to the bias functions of TNG300-1 simulations. Bias
parameters of HR5 simulations are almost identical to bias parameters
of TNG300-1 galaxies: HR5 mass limits $M_\star=1,~3,~10$ (in units
$10^9~M_\odot$) corresponds to magnitude $M_r$ limits of TNG300-1
simulation $M_r=-17.9,-19.6,-22.1$, for all simulation epochs with a
small scatter of the order $\pm 0.1$ mag.  Despite the different simulation programs and recipes of
galaxy formation and evolution, this similarity is remarkable.

Another similarity is in the shape of
correlation and bias functions for TNG100 and TNG300 simulations. At
epoch $z=0$, these functions for both simulations are very close; at
higher redshifts, the values of the bias functions of the simulation TNG300 for
luminous galaxies are higher; see Figs.~\ref{fig:Fig5} -- \ref{fig:Fig7}. This difference
is due to slightly various evolutionary stages of luminous galaxies in
TNG100 and TNG300 simulations, also seen in the shape of the luminosity
distributions in Fig.~\ref{fig:Fig3}.

Bias functions of TNG300-3 DM simulations are presented in the third
row of Fig.~\ref{fig:Fig5}, and of $\Lambda$CDM simulations in Fig.~5
of \citet{Einasto:2023aa}.  This comparison shows that in the
separation interval $1 \le r \le 100~\Mpc$ the shapes of bias
functions of TNG300-3 DM simulations are very close to the shapes of
bias functions of $\Lambda$CDM simulation in box of size
$L=256~\Mpc$. Both simulations have approximately equal volumes.  Also
the evolution of the bias parameters with cosmic epoch $z$ are
similar: compare bottom row of Fig~\ref{fig:Fig6} and left panels of
Fig.~6 by \citet{Einasto:2023aa}.  In both Figures at low particle
density limits $b(z,\rho_0)$ lines are almost independent on $z$, and
increase with $z$ for higher particle density limits $\rho_0$.

\subsubsection{The shape of correlation and bias functions on small separations}

At around separation $r \le 5~\Mpc$, the correlation and bias functions
depend on the structure of the halos.  
Galaxies and DM particles are located in identical halos, and the density-limited--DM particle pair counts are higher, which raises the amplitude of CFs.
This remarkable feature is well known from earlier studies, see among others
\citet{Chiang:2013tq}, \citet{Springel:2018aa} and
\citet{Einasto:2023aa}; it corresponds to the one-halo term in Halo Models
\citep{Asgari:2023aa}.  Halos are gravitationally stable
systems detached from the expansion.
In comoving units, halos were larger in the past. 

On small separations $0.1 \le r \le 2~\Mpc$, an important difference in bias functions between
galaxy and DM simulations emerges on Fig.~\ref{fig:Fig5};
compare top and bottom rows.  The bias functions of
simulations TNG100-1 and TNG300-1 are made for galaxy--galaxy pairs,
simulations TNG100-3 and TNG300-3 for particle density-limited DM
particle pairs.  In galaxy--galaxy pairs, the bias functions in this
separation interval have depressions, but bias functions of DM
particle pairs are flat, forming almost parallel lines for
particle density limits $\log\rho_0 \le -6.8$.
This depression in galaxy--galaxy pairs is also seen in the analysis by
\citet{Springel:2018aa}. These separations correspond to central
regions of halos (clusters of galaxies).  A possible explanation of this phenomenon is that near the centres of clusters, a large fraction of faint galaxies is ``eaten'' by more massive
galaxies.


\subsection{Amplitudes of correlation and bias functions}

\subsubsection{Bias parameters of low luminosity galaxies}

The basic difference between TNG100-1 and TNG300-1 galaxy simulations
on the one side and TNG100-3 and TNG300-3 DM particle simulations on
the other side is in the shape of bias functions in low and
intermediate luminosity $M_r$ and particle density $\log\rho_0$
ranges. As shown in Fig.~\ref{fig:Fig7}, the bias parameter $b(M_r)$
of galaxy simulations TNG100-1 and TNG300-1 is constant for a broad
interval of luminosity, $M_r \ge -20$ for present epoch $z=0$, and for
$M_r \ge -15$ for early epoch $z=5$. In contrast, the bias parameter
$b(\log\rho_0)$ of DM TNG100-3 and TNG300-3 simulations rises
continuously with increasing particle density limit $\log\rho_0$.  The
approximating of the function $b(M_r)$ to a low asymptotic level $b_0$
was
found  by \citet{Norberg:2001aa} from the 2dF Galaxy Redshift survey
and by \citet{Zehavi:2011aa} and \citet{Einasto:2020aa} for SDSS
galaxies.  A similar phenomenon was found by \citet{Einasto:2020aa}
for Millennium simulation galaxies in magnitude interval
$-17.4 \ge M_r \ge -20$, and for EAGLE simulations in magnitude
interval $-15.5 \ge M_r \ge -18.0$.

These differences between galaxy and DM simulations mean that the
transition of the DM filamentary web from higher to lower $\log\rho_0$
levels is a continuous process, but the transition of the
filamentary web of galaxies to lower luminosities has a sharp limit.
In other words, there is no population of dwarf galaxies in faint
DM filaments -- faint dwarf galaxies are located in the same
filamentary web as brighter ones
\ME{
and the properties of galaxies are largely shaped by their birthplace
in the cosmic web (initial conditions for galaxy formation)
\citep{Repp:2019ti, Einasto:2022aa}.}
Differences between filamentary webs
of galaxies and DM are clear in Fig.~\ref{fig:Fig2}. 
The upper panel displays the faint DM web that is absent from the bottom
panel of the galaxy-defined web; see also
\citet{Springel:2005aa, Springel:2018aa}.

The constant level of the bias function $b(M_r)$ at low luminosities
raises the question of the galaxy distribution in voids.  When
large voids were detected by \citet{Gregory:1978},
\citet{Joeveer:1978dz}, and \citet{Kirshner:1981a}, then 
\citet{Dekel:1986aa} and  \citet{Dekel:1986ab}  assumed that giant galaxies form in
high-density regions, but dwarf galaxies can also form in voids.  To
check the presence of void galaxies \citet{Einasto:1988aa,
  Einasto:1990aa} compared distributions of faint and bright galaxies
in and around the Virgo supercluster and found that both types of
galaxies occupy identical regions.  The study by
\citet{Lindner:1995ui, Lindner:1996tu} showed that dwarf galaxies are located near void boundaries, and that they are not  randomly distributed in voids. The void phenomenon was
studied by \citet{Peebles:2001kl}, \citet{Tinker:2009bh}, and
\citet{Neyrinck:2014qf}. Using numerical simulations, 
\citet{Tinker:2009bh} found that the boundary between filaments and
voids in the galaxy distribution is nearly as sharp for dwarfs as for
$\sim L_\star$ galaxies.  Note that this observation does not exclude the presence of some
isolated dwarf galaxies in the outer surrounding of brighter galaxies,
similar to dwarf galaxies observed recently by \citet{Rizzi:2017aa},
\citet{Karachentsev:2023aa} and \citet{Makarova:2023aa}.

For the present epoch $z=0$, the galaxy simulations TNG100-1 and TNG300-1
yield samples with almost identical asymptotic values of the
bias parameter of low-luminosity galaxies, $b_0=1.045$ and $b_0=1.044$
respectively.  This suggests that the mean asymptotic bias parameter of
the lowest luminosity galaxies in the present epoch has a mean value
$b_0=1.045 \pm 0.01$.  For the epoch $z=0$, the DM simulations TNG100-3 and
TNG300-3 also yield particle density-limited samples with similar bias
values, $b_0=1.195$ and $b_0=1.299$, respectively, with a mean value
$b_0=1.25 \pm 0.05$. The error was estimated from the difference of
$b_0$ values for simulations TNG100-3 and TNG300-3.  This bias value
is only 1.2 times higher than for galaxy samples.
For earlier epochs $b_0(z)$ curves for simulations TNG100 and TNG300
diverse, both for galaxy and particle-density-limited samples.  As discussed by
\citet{Einasto:2023aa}, in earlier epochs, fixed luminosity and
particle density limits correspond to more advanced stages of
evolution, see Fig.~\ref{fig:Fig3}, which raises the amplitudes of the bias
parameters. 

\subsubsection{Amplitudes of correlation and bias functions as
  cosmological parameters}

One possibility to define the bias parameter is to use Halo Models (HM) of large-scale structure, for a recent review, see \citet{Asgari:2023aa}.
In the HM model the bias is defined as a function of  mass, $b(M)$, and it satisfies   normalisation conditions:
$\int_0^\infty Mn(M) \dd{M} = \overline{\rho}$ and 
$\int_0^\infty Mb(M)n(M) \dd{M} =\overline{\rho}$, where $n(M)$ is the halo mass function,
and $\overline{\rho}$ is the mean comoving cosmological matter
density.  In the HM prescription, low-mass halos are anti-biased
($0 < b(M) <1$) with a constant asymptotic value at low mass.  In HM, the bias function is defined with respect to the characteristic
mass $M_\star$, where the transition to biased objects occurs.  HM is based on tacit or implicit assumptions that all matter is
contained in halos, and the matter in low-density regions outside
halos can be ignored. We drop the second assumption when we study the effect of
particles in low-density regions on the bias phenomenon.

It is well-known that amplitudes of correlation and bias functions
depend on cosmological factors: (i) cosmological parameters:
matter-energy densities $\Omega_b,~\Omega_m,~\Omega_{\Lambda}$, (ii)
the present {\em rms} matter fluctuation amplitude averaged over a
sphere of radius $8~\Mpc$, $\sigma_8$; (iii) luminosities of galaxies
\citep{Kaiser:1984}; (iv) systematic motions of galaxies in clusters
-- the finger of God effect,  (v) the flow of galaxies toward
attractors \citep{Kaiser:1987aa} and (vi) the thickness of observational samples, if 
3D CFs are determined by the inversion of 2D CFs \citep{Einasto:2021ti}. The present study is based on
numerical simulations of the cosmic web, where the last three effects are
not present.  We use simulations with fixed density and $\sigma_8$
parameters, thus the essential  cosmological parameter is the
luminosity of galaxies (and the particle density limit in DM
simulations).  However, our study shows that amplitudes of correlation
and bias functions depend on one more factor -- the fraction of matter
in voids and in the clustered population, which  is the topic of the
next subsection.

\subsubsection{Bias parameter and the fraction of matter in the
clustered population}

As noted in the Introduction, \citet{Einasto:1994aa, Einasto:1999ku,
  Einasto:2019aa, Einasto:2023aa} investigated the relation between
clustered and total matter using DM only numerical simulations of the
evolution of the cosmic web.  Clustered matter was identified with
samples of DM particles with local densities above a certain
threshold, $\rho \ge \rho_0$.  The main result of these studies was
the establishment of a relation between the bias parameter $b$ and the
fraction of matter in the clustered population, $F_c$: $b=1/F_c$.
\citet{Einasto:2023aa} found that the relation $b=1/F_c$ is well
fulfilled for low particle density limits $\rho_0$.  

Our analysis shows that correlation and bias functions depend on the
nature of objects used in their determination.  The expected
relationship between the bias parameter $b$ and the fraction of matter
in the clustered population $F_c$ is fulfilled in DM simulations
TNG100-3 and TNG300-3.  In these cases, the bias is defined using DM
particles: CFs based on numbers of particles in high-density regions
above the threshold level $\rho\ge \rho_0$, are divided by CFs found from
numbers of all particles of the full DM particle sample.
In simulations TNG100-1 and TNG300-1 the bias parameter is defined
using CFs of galaxies of luminosity $M_r$, divided by CFs of DM
particle samples.  In this case, in Eq.~(\ref{bias}), numerators and
denominators are objects of different nature: galaxies do not have a simple relationship to a threshold in the corresponding DM field.  Our analysis has shown
that the relationship between the bias parameter,
$b(M_r)$, and the respective fraction of the clustered matter, $F_c$,
is not fulfilled in this case.  Measured functions $b_0(z)$ for low-luminosity
galaxies of simulations TNG100-1 and TNG300-1 lie considerably lower
than expected functions $b_0(z)=1/F_c(z)$, by a factor of
$\approx 1.3$, over the whole range of simulation epochs $z$.

In the present study, we used the separation $r_0=10~\Mpc$ to measure
the bias parameter $b$.  Using this separation, we found for the bias
parameter of lowest luminosity galaxies, $b(z=0)=1.045$, both for
TNG100-1 and TNG300-1.  If a lower separation would be used,
$r_0 \approx 1~\Mpc$, then the bias parameter of lowest luminosity
galaxies at the present epoch would be smaller, $b(r) \le 1$, see left
panels of top and second rows of Fig.~\ref{fig:Fig5}. 
\citet{Springel:2005aa} has found that at the present epoch, galaxies
are slightly antibiased with $b =0.9$.  This means there is no
room for particles in the unclustered population, $F_v=1 - F_c$, if
the relation $b_c=1/F_c$ is valid.

During evolution, matter flows from low-density regions to
high-density ones. The outflow of matter from voids toward
superclusters was investigated using velocity field data
by \citet{Tully:2008ly, Tully:2014}, \citet{Carlesi:2016ul},
\citet{Sorce:2016ve}, \citet{Rizzi:2017aa}, and \citet{Anand:2019aa}.
However, gravity cannot evacuate voids completely, thus there is
always some matter in voids. \IS{Note that this matter has a low correlation compared to the highly clustered regions \citep{Repp:2022aa}. Therefore we can neglect its correlations to zeroth order and refer to it as unclustered.} Thus, the bias
parameter of galaxies must be greater than unity, over the whole range
of evolution epochs.

The principal result of this study is the establishment of the difference between properties of correlation
and bias functions of galaxies and DM particle samples.  In earlier studies, bias properties were
investigated using either galaxy or DM data: the fraction
of matter in the clustered population was not used.  For the first time, 
we use both types of test particles, galaxies and
samples of DM particles, in identical simulations.
Thus we avoid potential misinterpretations resulting from using different test particles and
data samples.

All previous studies have shown that the bias parameter depends on the
luminosity of galaxies (particle density limit).  However, only the
bias function of DM particle samples satisfies the $b=1/F_c$
criterion. This means that correlation and bias functions of only DM particle  samples \IS{properly} measure the relationship between galaxies and DM. The
general conclusion of this study is that correlation and bias
functions of galaxies and DM particle samples measure different
properties of the cosmic web.  This result is no surprise. As shown
recently by \citet{Ouellette:2023aa}, a topological analysis of
TNG simulations reveals the presence of differences between simulated
galaxies and DM halos.

\section{Conclusions}

We investigated properties of correlation and bias functions and bias
parameters using two types of data: simulated galaxies and DM
particles.  For both data types, we applied several input sources, for
galaxies TNG100-1, TNG300-1 and HR5 simulations, and for DM particle
samples TNG100-3, TNG300-3 simulations and $\Lambda$CDM simulations by
\citet{Einasto:2023aa}.  Our analysis showed that essential properties
of correlation and bias functions of simulated galaxies using various
data sources are consistent. The same consistency of correlation and
bias functions is also observed using all particle density-limited
simulations.  In contrast, essential differences exist between
correlation and bias functions of luminosity-limited samples of
galaxies on the one side and particle density-limited samples on the
other. The fundamental results of our study can be listed as follows.

\begin{enumerate}
  
\item{} The bias parameter of low luminosity galaxies approaches an
  asymptotic level $b(M_r) \rightarrow b_0$; at the present epoch
  $b_0=1.045 \pm 0.01$ for galaxy samples in TNG100-1 and TNG300-1 and
  HR5 simulations.  A flat region of the bias function $b(M_r)$ at
  low luminosities suggests that faint dwarf galaxies are located in
  the same filamentary web  as brighter galaxies.

\item{} The bias parameters $b(\rho_0)$ of particle density $\rho_0$
  limited samples of DM particles of TNG100-3 and TNG300-3 simulations
  form a continuous sequence with decreasing $\rho_0$, which suggests
  that the transition of the filamentary DM web from higher to lower
  particle density limit $\rho_0$ is continuous.  For the present
  epoch, the DM simulations TNG100-3 and TNG300-3 yield for the bias
  parameter a value $b_0=1.25 \pm 0.05$. This should correspond to the lowest luminosity galaxies.
  
\item{} Cosmic web consists of filamentary structures of various
  densities. The fractions of matter in the clustered and
  the unclustered populations are less than unity. For this reason,
  the bias parameter of the clustered matter is $b>1$ for all cosmic epochs.
  
\item{} Bias parameter $b_0$ from density-limited DM samples in TNG100-3 and TNG300-3 agrees with expectations from the fraction of particles in the clustered population: $b_0=1/F_c$.

\item{} Bias parameter $b_0$ of galaxy samples in TNG100-1 and
  TNG300-1 is in disagreement with the $b_0=1/F_c$ constraint. This
  means that correlation and bias functions, calculated for objects of
  different natures (galaxies vs. density-limited samples of DM
  particles) describe properties of the cosmic web differently.
  
\end{enumerate}
  
We do not fully understand the differences in correlation and bias functions for galaxies and density-limited DM populations. In particular, the measurements conflict with predictions from the fraction of particles in voids.  We note that  opinions on the role of the unclustered
matter in voids to general properties of the web are different.
\citet{Einasto:1994aa,Einasto:2023aa} used a simple analytic model to
estimate the fraction of matter in voids. Their analysis suggests that
the void fraction decreases with time.  In contrast, the
analytic model by \citet{Sheth:2004ly} indicates that the fraction of
mass in voids is constant over time.  The role unclustered matter in voids plays in the clustering of the cosmic web deserves further study.

\section*{Acknowledgements}

We thank the TNG  collaboration for publicly releasing
their simulation data and example analysis scripts, 
Neta Bahcall, Jim Peebles, Dmitri Pogosyan and Brent Tully for stimulating discussions, and the anonymous referee for useful suggestions.

This work was supported by institutional research funding IUT40-2 of
the Estonian Ministry of Education and Research, by the Estonian
Research Council grant PRG803, and by Mobilitas Plus grant MOBTT5. We
acknowledge the support by the Centre of Excellence ``Dark side of the
Universe'' (TK133) financed by the European Union through the European
Regional Development Fund. The study has been also supported by Kavli
Institute for Theoretical Physics, University of California, Santa
Barbara, through the program ``The Cosmic Web: Connecting Galaxies to
Cosmology at High and Low Redshifts'', and by ICRAnet through a
professorship for Jaan Einasto.
 
\section*{Data availability}

All data on TNG100 and TNG300 simulations used in this work are 
publicly available.  Any other data will be shared
upon reasonable request to the corresponding author.

\bibstyle{mnras}
\bibliographystyle{mnras} 

\bsp	
\label{lastpage}

\end{document}